\documentstyle[aps,multicol,amssymb,graphicx]{revtex}
\begin{document}
\bibliographystyle{apsrev}
\title{Diffraction of complex molecules by structures made of light}
\author{Olaf Nairz, Bj{\"o}rn Brezger, Markus Arndt, Anton Zeilinger}
\address{Universit\"at Wien, Institut f\"ur Experimentalphysik, Boltzmanngasse 5, A-1090 Wien, Austria}
\maketitle
\begin{abstract}
We demonstrate that structures made of light can be used to
coherently control the motion of complex molecules. In particular,
we show diffraction of the fullerenes C$_{60}$ and C$_{70}$ at a
thin grating based on a standing light wave. We prove
experimentally that the principles of this effect, well known from
atom optics, can be successfully extended to massive and large
molecules which are internally in a thermodynamic mixed state and
which do not exhibit narrow optical resonances. Our results will
be important for the observation of quantum interference with even
larger and more complex objects.
\end{abstract}

\begin{multicols}{2}
The great success of atom optics has stimulated the question
whether it is possible to extend the methods developed in this field
to more complex and massive quantum objects.
In this letter we demonstrate for the first time the coherent control
of the molecular motion using optical structures for macromolecules.

One possible manipulation technique for molecules has been
demonstrated earlier with the use of material nano\-struc\-tures.
They have for example successfully been used in atom
interferometry \cite{Pritchard1991}, in the determination of the
bond length of a helium dimer \cite{Grisenti2000} and more
recently in interference experiments with the fullerenes
C$_{60}$\cite{Arndt1999}  and C$_{70}$\cite{Nairz2000}.  The use
of solid nano\-structures has the advantage of being universal and
largely independent of the detailed internal character of the
diffracted object.  However, since the structure dimensions have
to be of the order of 100\,nm, material devices are extremely
fragile and can be blocked or destroyed by the molecules.

In contrast to that, diffraction structures made of
light are promising alternatives:
The periodicity can be perfect, the transmission is high
and one can realize different types of
gratings. Light may act as a
real or imaginary index of refraction for matter waves and thus form
a phase grating or an absorption grating.

\begin{figure}[ht]
\begin{center}
\includegraphics[width=1.00\columnwidth]{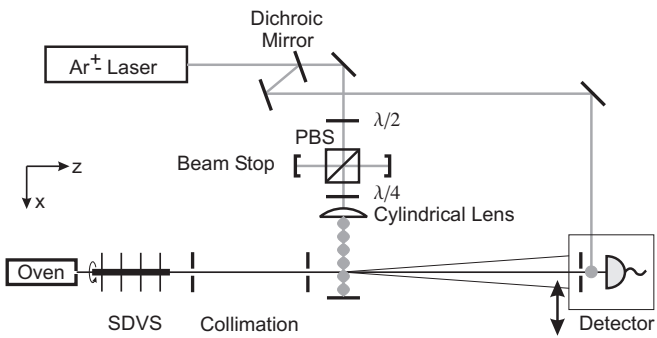}
\end{center}
\caption{An oven generates a fullerene beam. Collimation and
velocity selection prepare the required transverse and
longitudinal coherence. The beam interacts with a circularly
polarized standing light wave of periodicity 257 nm. The power is
variable between 0 and 9.5 W and focused to $w_z \times w_y =0.05
\times 1.3$ mm$^2$. The molecular diffraction pattern after the
light grating is recorded by a scanning optical ionization
stage.\label{setup}}
\end{figure}

For atoms amplitude gratings can be based on various effects which
lead to an effective spatially periodic depletion  of relevant
states of the atomic beam \cite{Abfalterer1997,Kunze1997}. For
example, the extraction of particles using ionization in a
standing wave is conceivable.

Phase gratings based on the non-dissipative dipole force were
demonstrated for the diffraction of atomic beams both in the
thin-grating or Raman-Nath regime \cite{Moskowitz1983} and in the
thick-grating or Bragg regime \cite{Martin1988}. They have been
successfully implemented to build up complete Mach-Zehnder
interferometers \cite{Rasel1995,Giltner1995} and they find
applications in atom lithography \cite{Timp1992,Schulze2001} or
the manipulation of Bose Einstein condensates \cite{Deng1999}.
Combinations of absorptive and phase structures can also be used
for complex blazed gratings\cite{Keller1997}.

In spite of the great success in atom optics, light gratings have
not yet been applied to larger molecules. This is mainly due to
significant differences between atoms and molecules which have to
be taken into account in a practical realization: The optical line
widths of large molecules are typically of the order of 100\,THz
instead of a few MHz in the atomic regime. In the case of
fullerenes the electrical polarizability, mediating the coupling
to the optical field, remains constant within a factor of two
throughout the whole frequency range from DC to UV. The high
complexity of these molecules leads to non-negligible absorption
from the ultra-violet well into the visible wave length region and
for the realization of a pure phase grating one would think that
absorption should be excluded. This reasoning is based on the
experience  from atom optics that absorption is usually followed
by spontaneous emission, which carries which-path information into
the environment.

In the following we will demonstrate that the principles of light
gratings can actually be successfully carried over to fullerenes,
which are internally in a thermodynamic mixed state. Even in the
presence of absorption the fullerenes may maintain coherence between states of
equal energy since the absorbed quanta are trapped in the
molecules.

A schematic of the setup is shown in fig.~\ref{setup}. Essential
parts of the setup are already described in \cite{Arndt1999} and
an in depth characterization of the detector is given in
\cite{Nairz2000}. A ceramic oven containing the fullerene powder
is heated to approximately 900~K. The fullerene beam exits with a
most probable velocity of 200\,m/s and 170\,m/s in the case of
C$_{60}$ and C$_{70}$, respectively, both with a FWHM velocity
spread of $\bigtriangleup v/v \approx 0.6$. In order to separate
the individual diffraction peaks in our experiment we select only
a slow part of the velocity distribution using a slotted disc
velocity selector (SDVS)\cite{Scoles}. We select a most probable
velocity of approximately 120~m/s with a velocity spread of
$\bigtriangleup v/v \approx 0.17$, which allows to keep about
6\,\% of the initial molecular flux. The beam is collimated by two
vertical slits of $7\,\mu$m and $5\,\mu$m separated by 1.13\,m.

Both the standing light wave and the detection are implemented
using the same 27~W, multi-line visible Ar$^{+}$-laser.  Using a
dichroic mirror with a transmission of $T>76\,\%$ at
$\lambda_{\mathrm{L}}=514.5$\,nm  and $T<1\,\%$ for all other
laser lines we isolate the single green line  for the standing
wave from the color mixture used in the ionization. The green
power can be varied continuously up to 9.5\,W by rotating a
halfwave-plate in front of a polarizing beam splitter (PBS). The
quarterwave-plate between the polarizing beam splitter and the
mirror is set to $45^{\circ}$ so that the back-reflected light is
deflected by the PBS and cannot return into the laser. The
ensemble averaged molecule-light interaction is expected to be
insensitive to the light polarization since the molecules enter in
a thermal mixture of rotational and vibrational states.

In order to ensure that the light potential seen by the fullerenes
does not average out during their transit through the standing
light wave, the nodal planes of the standing light wave have to be
parallel to the molecular beam.  This is facilitated by focusing
the laser beam in the direction of the molecular trajectory. The
cylindrical lens has a focal length of 30~cm and reduces the laser
beam from a $1/e^2$-radius of 1.3~mm to $w_z \approx 50~\mu $m at
the mirror surface. The final alignment of the mirror could be
achieved with the collimated fullerene beam itself. The mirror was
first moved into the fullerene beam, so that it cut a part of it,
and was then rotated around the axis centered on the mirror
surface and normal to both the fullerene beam and the laser beam.
A symmetric decrease in count rate when rotating the mirror in
both directions indicated the correct alignment. After this
procedure we estimate the mirror to be parallel to the molecular
beam by better than 0.5~mrad.  During the experiment the beam
passed the mirror surface in a distance of 100\,$\mu$m.

The height of the detected fullerene beam was determined using a
knife edge and amounted to $(625 \pm 70)\,\mu$m FWHM. A coarse
overlap between the molecular beam and the standing wave was
achieved using a horizontal slit. The final alignment was done
using the diffraction pattern itself: maximal diffraction
efficiency indicated perfect overlap.

In order to record the diffraction pattern 1.2 m downstream we
scanned the focused detection laser beam (power 17\,W, beam waist
$\approx 4 \,\mu $m) across the molecular beam in $2 \,\mu $m
steps. The fullerenes were ionized by absorption of 20 to 30 blue
and green photons and the ions were subsequently counted. The
width of the detector at this power for C$_{60}$ is approximately
$6 \,\mu $m \cite{Nairz2000}. For C$_{70}$ the absorption cross
section is much higher than for C$_{60}$ and so the width of the
detector would be larger. In order to ensure a comparable detector
width for both types of fullerenes, we placed a $5 \,\mu $m
precision slit in front of the laser beam and moved it together
with the focusing optics.

\begin{figure}[th!]
\begin{center}
\includegraphics[width=1.00\columnwidth]{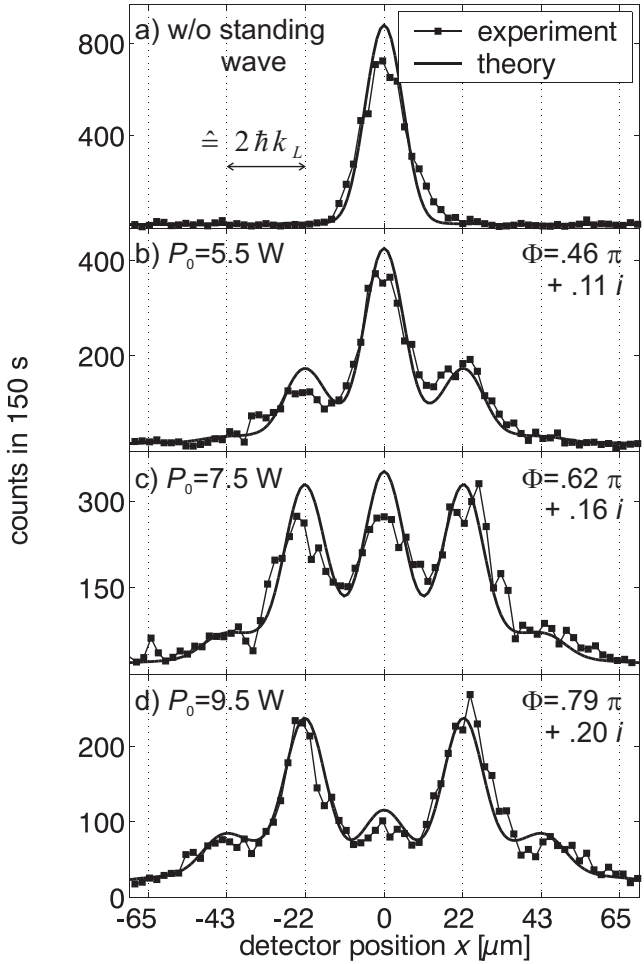}
\end{center}
\caption{Interference patterns for C$_{60}$ for different laser
powers. $\Phi$ is the phase shift parameter as defined in
eq.~\ref{eq:phidef}. Twice its imaginary part gives the mean
absorbed photon number (eq.~\ref{eq:nbarexp}). The diffraction
efficiency into each of the first diffraction orders in case d can
be estimated to be about 25~\%. \label{fig60}}
\end{figure}

\begin{figure}[ht]
\begin{center}
\includegraphics[width=1.00\columnwidth]{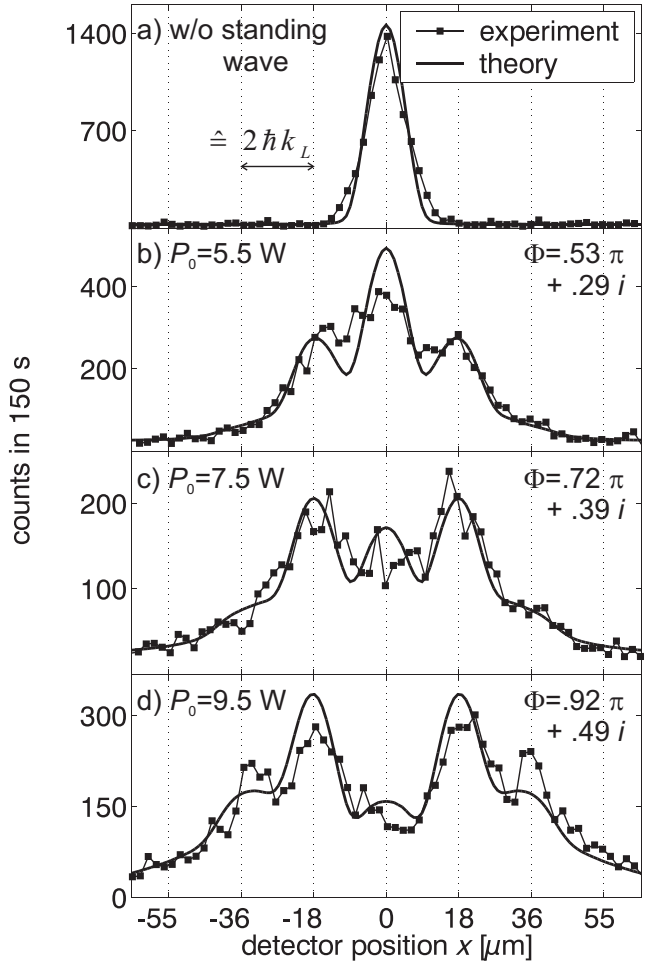}
\end{center}
\caption{Interference patterns for C$_{70}$ for the same laser
powers as in fig.~\ref{fig60}.\label{fig70}}
\end{figure}

Since the whole vacuum apparatus drifted by several micrometers
within a few hours we had to limit the recording time per
diffraction pattern to less than an hour. In order to obtain a
sufficient counting statistics we performed several (typically 15)
runs. Within each run we recorded both an undiffracted and a
diffracted picture by blocking and unblocking the standing light
wave for each individual detector position. The reference pictures
allowed us to center the diffraction patterns before they were
summed. The far field diffraction patterns were recorded at
different laser powers both for C$_{60}$ and C$_{70}$ as shown as
full squares in figs.~\ref{fig60} and~\ref{fig70}.

The continuous curves in these graphs are the result of our
theoretical model: The fullerenes do not have a permanent electric
dipole moment but a finite polarizability $\alpha$. The electric
field $E$ of the laser induces in first order a dipole moment and
in second order a potential
$V^{\mathrm{dip}}(x,z)=-\frac{1}{2}\alpha \overline{E^2(x,z,t)} =
-\alpha I (x,z)/(2c \varepsilon_0)$, proportional to the local
intensity $I(x,z)$.

The action of the standing light field on the molecular de Broglie
waves is described as that of a thin grating in the Raman-Nath
regime \cite{Moskowitz1983}: the displacement of the molecules due
to diffraction and within the light field itself is much smaller
than one  grating period. Immediately behind the grating the
spatial distribution of the molecules is therefore essentially
unchanged. We thus obtain the following transmission function
\cite{atomoptik} for those molecules that leave the interaction
zone in the same internal state as before:

\begin{eqnarray}
\label{eq:zerotrans}
t^{\mathrm{dip}}(x)&=&
 \exp\left(-\frac{i}{\hbar}\int V(x,z(t))\,\mathrm{d}t\right)\\
\label{eq:zerotranscos}
 &=&
 \exp\left(2i\Phi\, \cos^2\left(k_{\mathrm{L}}
 x\right)\right)
\end{eqnarray}
 where the phase shift has a mean
 value of
\begin{equation}
\label{eq:phidef}
 \Phi =\sqrt{\frac{2}{\pi}}\,\frac{P_0\,\alpha}{w_y v_z \hbar c
 \varepsilon_0}.
\end{equation}

Eq.~\ref{eq:phidef} holds for the passage of the molecules through
the center of an elliptical Gaussian laser profile of vertical
$1/e^2$ width $w_y$ as in the experiment. $P_0$ is the power of
each constituent running wave and $v_z$ the molecule velocity. By
virtue of the grating periodicity in $x$ with
$\lambda_{\mathrm{L}}/2$, far-field peaks deflected by $\Delta p_x
= \pm m\cdot 2\hbar k_{\mathrm{L}}$ ($m \in{\mathbb{N}}$) result.
Their relative intensities are calculated by Fourier decomposition
of eq.~\ref{eq:zerotranscos} and  are given by the squared Bessel
functions $J_m^2\left(\Phi\right)$. In particular the zeroth
diffraction order can be suppressed for $J_0(\Phi)=0$ --- in
contrast to diffraction at a mechanical grating. The experimental
parameters for fig.~\ref{fig60}\,d are close to this situation.

So far we have neglected the imaginary part of the polarizability
$\alpha$ which describes the absorption of grating photons by the
molecules with a cross section
$\sigma={\mathrm{Im}}(\alpha){k_{\mathrm{L}}}/{\varepsilon_0}$.
Absorption processes change both the internal state and the
momentum of the molecule but lead neither to a loss of signal nor
to spontaneous emissions.

Those molecules that absorb one or more photons will change to a
different internal state  but still contribute to the detected
signal. We take this into account by introducing transmission
functions $t_{0\to n}(x)=t^{\mathrm{dip}}(x)\cdot t_{0\to
n}^{\mathrm{abs}}(x)$ and an incoherent sum over the contributions
for different absorbed photon numbers $n$. The dipole part
$t^{\mathrm{dip}}(x)$ is taken to be the same for all $n$, where
we replace $\alpha$ with ${\mathrm{Re}}\left(\alpha\right)$ in
eq.~\ref{eq:phidef}. The absorptive part follows from assuming a
Poissonian statistics $p_{{\bar{n}}(x)}(n)$ with a mean absorbed
photon number ${{\bar{n}}(x)}$:
\begin{eqnarray}
\label{eq:nbardef}
 {\bar{n}}(x) &=& \frac{\sigma}{\hbar \omega_{\mathrm{L}}} \int I(x,z(t))\,\mathrm{d}t
\\
 \label{eq:nbarexp}
 &=& {\mathrm{Im}}(\Phi)\,4 \cos^2\left(k_{\mathrm{L}} x\right)
\\
 \label{eq:ntrans}
  t_{0\to n}^{\mathrm{abs}}(x) &=& \sqrt{p_{{\bar{n}}(x)}(n)}
\left(\frac{E(x,z)}{\left|E(x,z)\right|}\right)^n
\end{eqnarray}

The use of Poissonian statistics is equivalent with assuming that
the absorption processes are independent. This is justified by the
--- compared to the inverse absorption rate --- fast internal
relaxation processes. The first factor in eq.~\ref{eq:ntrans}
conserves the number of molecules since $\sum_{n=0}^\infty
\left|t_{0\to n}^{\mathrm{abs}}(x)\right|^2 = 1$. The second
factor, representing a phase independent of $z$, stems from
momentum conservation: The molecules inherit the phase of the
absorbed photons. In effect, the transmission function $t_{0\to
n}(x)$ contains only Fourier components corresponding to odd
(even) multiples of $\hbar k_{\mathrm{L}}$ for odd (even) $n$. In
our experiment the resolution is good enough to clearly resolve a
peak spacing of $2\hbar k_{\mathrm{L}}$. Peaks at odd multiples of
$\hbar k_{\mathrm{L}}$ fill in the minima and decrease the
contrast of the interference pattern.

The simulation is based on the described transmission functions
followed by free-space propagation and it contains no free
parameters. An incoherent sum is performed over source points in
the first slit, over the measured longitudinal velocity
distribution and over the measured vertical fullerene distribution
in the plane of the grating. We take into account the vertical
laser profile and the finite detector resolution. The presented
formalism relies on the assumption that all populated molecular
states have the same complex polarizability and that all internal
states are detected with the same efficiency. The complex ground
state polarizabilities at 514 nm which enter our model are
$\alpha_{\mathrm{C60}}=(101+8i)$\AA$^3\cdot 4\pi\varepsilon_0$
\cite{Sohmen1992} and $\alpha_{\mathrm{C70}}=(118+20i)$\AA$^3\cdot
4\pi\varepsilon_0$ \cite{Andersen2000}. The imaginary parts
correspond to absorption cross sections of $\sigma_{C60}=1.2\times
10^{-17}\,$cm$^2$ and $\sigma_{C70}=3.1\times 10^{-17}\,$cm$^2$.

Both fig.~\ref{fig60} and fig.~\ref{fig70} show a very good
overall agreement between the theoretical model and our
experimental results. Some broadening in the experimental curves
is attributed to non-perfect angular alignment of collimation and
detection slits and similar imperfections. Only in the case of
C$_{70}$ interacting with the maximum laser power
(fig.~\ref{fig70}~d) we see a surprising deviation between the
experiment and our model. The diffracted intensity at $\pm 4\hbar
k_{\mathrm{L}}$ seems to be higher in the experiment than in the
model. The significant difference of C$_{70}$ as compared to
C$_{60}$ is its higher absorption cross section for green light
--- therefore in the simulation at maximum laser power 12\,\% of
the C$_{70}$ molecules absorb two photons, thus also contributing
to the peaks at even multiples of $\hbar k_{\mathrm{L}}$, but only
4\,\% in the case of C$_{60}$. The effect could even be stronger
due to the higher absorption cross section \cite{Gratz1999} or a
higher polarizability of the excited states.

Concluding, we have shown that optical gratings for fullerenes
work very well using intense visible laser light. The diffraction
patterns can be fully understood when the standard theory for
dipole forces is supplemented  by the effect of photon absorption.
Optical gratings are generally important for quantum interference
experiments with even larger and more complex objects --- under
the general requirement of not too prominent absorption at the
employed laser wavelength --- since they have an excellent
regularity, transmission and efficiency and cannot be destroyed by
the diffracted particles. In particular, we have shown that a
C$_{60}$ beam can be split at will in two or three arms with good
efficiency (cf.\ fig.~\ref{fig60}\,c and d). This is important in
future applications of interferometers using beam splitters based
on light.

Optical gratings possess useful scaling properties: Mass and
polarizability have roughly the same scaling behaviour. They are
both proportional to the volume of the object. Conceptually, light
gratings may ultimately even be used for particles the size of
which is comparable to the grating constant. It will also be
interesting to see the influence of for instance the symmetry of
the molecules or their internal excitation on their ability to
interfere. Finally, one may envisage molecule holography with
dedicated light structures.

\begin{acknowledgments}
We acknowledge help in the setup of the experiment by Lucia
Hackerm\"{u}ller. This work has been supported by the European TMR
network, contract no. ERBFMRXCT960002, and by the Austrian Science
Foundation (FWF), within the project F1505. O.~N. acknowledges
financial support by the Austrian Academy of Sciences.
\end{acknowledgments}

\end{multicols}
\end{document}